\documentclass[11pt] {article}

\usepackage{amsmath, amsthm, amssymb}
\usepackage[dvips]{graphics}
\usepackage{hyperref}
\input xy
\xyoption {all}

\newtheorem{thm}{Theorem}[section]
\newtheorem{cor}[thm]{Corollary}
\newtheorem{lem}[thm]{Lemma}

\theoremstyle{definition}
\newtheorem{defn}[thm]{Definition}
\theoremstyle{remark}

\numberwithin{equation}{section}

\newcommand{\set}[1]{\left\{#1\right\}}

\newcommand{\R}{\mathbb{R}} 
\newcommand{\C}{\mathbb{C}} 
\newcommand{\Z}{\mathbb{Z}} 

\newcommand{\modd}{(\mbox{mod d})}
\newcommand{\gset}[1]{\left\langle #1 \right\rangle}
\newcommand{\ket}[1]{| #1 \rangle}
\newcommand{\bra}[1]{\langle #1 |}
\newcommand{\ip}[2]{\langle #1 | #2 \rangle}

\newcommand{\nzer}[2]{\mathcal{N}_{#1}(#2)}  

\newcommand{\smallGate}[1]{\left( \begin{smallmatrix} #1 \end{smallmatrix} \right)}

\newcommand{\pauli}[2]{\mathcal{P}_{#1}^{\otimes #2}}
\newcommand{\pauliOne}[1]{\mathcal{P}_{#1}}
\newcommand{\clifford}[2]{\mathcal{C}l_{#1}^{\otimes #2}}
\newcommand{\cliffordOne}[1]{\mathcal{C}l_{#1}}

\newcommand{\be}{\begin{equation}}
\newcommand{\ee}{\end{equation}}
\newcommand{\bea}{\begin{eqnarray}}
\newcommand{\eea}{\end{eqnarray}}
\newcommand{\bes}{\begin{equation*}}
\newcommand{\ees}{\end{equation*}}
\newcommand{\beas}{\begin{eqnarray*}}
\newcommand{\eeas}{\end{eqnarray*}}

\begin{document}

\title{Valence bond solid formalism for d-level one-way quantum computation\footnote{A preliminary version of this work was presented in a poster at the CNRS summer school on Quantum  Logic and Communication, Corsica August 2004.}}

\author{Sean Clark\footnote{Department of Computer Science, University of Bristol, Woodland Road, Bristol, BS8 1UB, England. email: sean.clark@bristol.ac.uk}}
\maketitle
\begin{abstract}
The d-level or qudit one-way quantum computer (d1WQC) is described using the valence bond solid formalism and the generalised Pauli group.  This formalism provides a transparent means of deriving measurement patterns for the implementation of quantum gates in the computational model.   We introduce a new universal set of qudit gates and use it to give a constructive proof of the universality of d1WQC.  We characterise the set of gates that can be performed in one parallel time step in this model.  
\end{abstract}

\setlength{\parindent}{0pt}
\setlength{\parskip}{1ex plus 0.5ex minus 0.2ex}

\section{Introduction}
Since its introduction the one-way quantum computer (1WQC) \cite{1wqc_orig, 1wqc} has sparked interest in areas including the study of resources for quantum computation, the complexity of algorithms \cite{comp_1wqc}, and practical implementation schemes for quantum computing \cite{optic_1wqc, fault_tolerant_1wqc, eff_lin_optic_1wqc, quantum_dot_1wqc, experimental1wqc}.

Comparing the 1WQC with the standard quantum circuit (QC) model allows us to ask questions about the resources required for quantum computation.  The standard QC model requires (at least in a perfect world) preparation of the zero state, controlled unitary evolution of a universal set of gates and measurement in the computational basis.  This compares to the 1WQC which requires preparation of a multipartite entangled cluster state and the ability to  perform measurements in classically computed adaptive bases.  There have also been comparisons \cite{unify_leung, unify_jorrand, unify_childs, vbs, jozsa_mqc} showing that the 1WQC is equivalent to another model of measurement based quantum computation known as teleportation-based quantum computation (TQC) \cite{tqc_orig_nielsen, tqc_orig_leung}.  The valence bond solid (VBS) formalism \cite{vbs} of the 1WQC provides a fundamental basis for such a comparison.

Here we extend the use of the VBS formalism to describe the workings of the 1WQC for d-level systems or qudits which is known as the d-level one-way quantum computer (d1WQC).  The d1WQC was first introduced in \cite{d1wqc} in which its workings are described in terms of an irreducible representation of Manin's quantum plane algebra \cite{manins_qpa}.  The VBS formalism provides a clear representation of the workings of the d1WQC and opens the way for a variety of natural generalisations.

The construction of the d1WQC given here exposes a special role of the group of generalised Clifford operations: quantum circuits of such operations, when implemented on the d1WQC can be performed in one parallel measurement time step followed by poly-logarithmic classical processing.  We give a full characterisation of the Clifford group of circuits for d-level systems in the appendix that differs from the more formal approach of \cite{cliffGenModArith}.

The workings of the qubit 1WQC were introduced in \cite{1wqc_orig, 1wqc} and a review of this and measurement-based quantum computation is given in \cite{jozsa_mqc}.  Computation in this model proceeds by producing a highly entangled state called a cluster state and then performing measurements on each of the qubits.  The cluster state is described by the local interactions between its constituent quantum systems. Each qubit in the cluster is measured during the computation using one-qubit projective measurements in a chosen basis that may be calculated classically from any previous measurement results. The specification of the cluster state and choice of basis for each of the measurements together define the algorithm performed.  The d1WQC is a natural extension of the 1WQC in which the constituent systems are d-level quantum systems or qudits.  We describe an arbitrary d-level cluster state constructed from the VBS picture using the generalised Pauli group of quantum gates which are defined in the next section.  We  also show how to perform a universal set of gates on multi-qudit systems in this model.

The paper proceeds as follows.  We start by defining the Pauli and Clifford groups for qubits and their generalisations to systems of qudits and prove a theorem characterising the Clifford group in prime dimension.   We proceed to give definitions of cluster states and a VBS states of qudits and show that a cluster state can be obtained from a VBS state by applying a suitable projector.  We then go on to describe the workings of teleportation-based quantum computation on VBS states by constructing a parameterised one-qudit gate and the two-qudit generalised controlled-Z gate.   We show how these constructions can be concatenated and prove that they allow for universal quantum computation.  Next we show how the same projector can be used to transparently derive measurement schemes for gate implementations on the d1WQC.  Finally we mention the implications of this formalism  for the parallel complexity of generalised Clifford circuits.

\section{The generalised Pauli group}\label{chapter_pauli}
A basic ingredient in our description of the workings of the d1WQC is the generalised Pauli group of quantum gates.  In this section we review the Pauli group for systems of qubits and describe the natural extension to systems of  qudits.  We also define the the Clifford group of gates that normalise the Pauli group.  In doing so we establish the notation used throughout the paper.

The Pauli group of quantum gates on one qubit, denoted $\pauliOne{2}$, is defined in terms of its generators $\sigma_x$ and $\sigma_z$.  $\pauliOne{2} = \gset{\sigma_x, \sigma_z}$ where
$\sigma_x = \smallGate{ 0 & 1 \\ 1 & 0}$, and $\sigma_z = \smallGate{ 1 & 0 \\ 0 & -1 }$.

We note that this differs from the usual definition which usually includes the gate $\sigma_y = \smallGate{ 0 & -i \\i & 0}$ amongst the generators.  We will refer to our definition as the real Pauli group and the more usual definition as the complex Pauli group.

We extend the Pauli group by tensor products, leading to the Pauli group on $n$-qubits, $\pauli{2}{n}$, such that
\be
\pauli{2}{n} = \left\{\bigotimes_{k=1}^{n} p_k : p_k \in \pauliOne{2} \right\}.
\ee
The normaliser $\nzer{U}{G}$ of any complex matrix group $G \subset U(d)$ within the unitary group $U(d)$ is defined to be
\be
\nzer{U(d)}{G} = \set{ N \in U(d): \forall A \in G\mbox{, } \exists A^\prime \in G \mbox{, } c \in \C  \mbox{ s.t. }NAN^\dagger = cA^\prime}.
\ee
Note that this differs from the standard mathematical definition in that we allow for an extra constant $c$ (which necessarily has unit modulus).

The Clifford group on $n$ qubits is defined as being the normaliser of the Pauli group within the unitary group.

\be
\clifford{2}{n} = \nzer{U(2^n)}{\pauli{2}{n}}.
\ee
The more general definition of normaliser is justified in the current context since both in teleportation and quantum computation we consider two elements of the Pauli group to be equivalent if they differ only by some global phase factor.  We also note that in using our definition the real and complex Pauli groups have the same normaliser whereas they do not given the standard definition.

Some further gates used in this paper are the following.
The one-qubit Hadamard gate is
\be
H = \frac{1}{\sqrt{2}}\smallGate{ 1 & 1 \\ 1 & -1 }.
\ee

The $\frac{\pi}{4}$ phase gate is
\be
 S = \left( \begin{smallmatrix} 1 & 0 \\
0 & i
\end{smallmatrix} \right).
\ee
The two-qubit controlled-NOT gate is
\be
 CNOT = \smallGate{ 1 & 0 & 0 & 0 \\
                                        0 & 1 & 0 & 0 \\
                                        0 & 0 & 0 & 1 \\
                                        0 & 0 & 1 & 0 }.
\ee
The controlled-Z gate is
\be
C_Z = diag(1,1,1,-1)
\ee
where $diag$ denotes a diagonal matrix with the given entries.

In fact $H$, $S$, $CNOT$ and $C_Z$ are all gates in the Clifford group and furthermore it was shown in \cite{gottesmanPhd} that, up to a global phase factor, $H$ and $S$ together generate $\cliffordOne{2}$ and $H$, $S$, and $CNOT$ together generate $\clifford{2}{n}$ for any $n$.

We now define the natural generalisation of the Pauli group to systems of qudits.

Let the one qudit gates $X$ and $Z$ be such that for $j \in \Z_d$ (where $\Z_d$ denotes the ring of integers modulo $d$ sometimes denoted $\Z/d\Z$)
\be
X\ket{j} = \ket{j+1\modd}
\ee
\be
Z\ket{j} = \omega^j\ket{j}
\ee
where $\omega = exp\left(\frac{2\pi i}{d}\right)$ is the $d^{th}$ root of unity.  We note the fundamental relation
\be
ZX = \omega XZ.
\ee

\begin{defn} \label{defn_gen_pauli}
The generalised Pauli group on one qudit, $\pauliOne{d} = \gset{X, Z} $, is defined to be the group generated by $X$ and $Z$, and the Pauli group on $n$ qudits is defined as
\begin{equation}
\pauli{d}{n} = \left\{\bigotimes_{k=1}^{n} p_k : p_k \in \pauliOne{d} \right\}.
\end{equation}
\end{defn}

Using the relation $ZX = \omega XZ$ we note that we can express $\pauli{d}{n}$ as
\be
\pauli{d}{n} = \left\{ \omega^kZ_1^{a_1}X_1^{b_1}...Z_n^{a_n}X_n^{b_n} : a_j,b_j,k \in \Z_d \right\}
\ee
Here the subscripts label upon which qudit the operator acts. Often we will not be interested in the the global phase $\omega^k$.  In this case we may considered the central quotient group $\pauli{d}{n} / Z(\pauli{d}{n})$ (where $Z(\pauli{d}{n})$ denotes the centre of the group) with representatives of the form $Z_1^{a_1}X_1^{b_1}...Z_n^{a_n}X_n^{b_n}$.

\begin{defn}\label{defn_gen_clifford}
The Clifford group on $n$ qudits, $\clifford{d}{n}$, is defined to be the normaliser of $\pauli{d}{n}$ in $U(d^n)$.  That is
\be
\clifford{d}{n} = \nzer{U(d^n)}{\pauli{d}{n}}.
\ee
\end{defn}

The generalisation to the qudit case of the $H$, $S$, controlled-NOT and controlled-Z gates are as follows. $H$ becomes the quantum Fourier transform on one qudit, which we denote by $F$.
\be
F \ket{j} = \frac{1}{\sqrt{d}}\sum_{m \in \Z_d} \omega^{jm} \ket{m}
\ee
For the case where $d$ is odd we have the definition
\be
S \ket{j} = \omega^{\frac{j}{2}(j+1)} \ket{j}
\ee
and
\be
C_X \ket{j}\ket{k} = \ket{j}\ket{j+k(\mbox{mod }d)} \mbox{, }\label{defn_cz}
C_Z \ket{j} \ket{k} = \omega^{jk} \ket{j}\ket{k}.
\ee
In the appendix we give a proof of the following theorem:

\begin{thm}
Any Clifford circuit on $n$ qudits, where $d$ is an odd prime, can be constructed as quantum circuit up to a global phase from the gates $\set{C_X,F,S}$.
\end{thm}

\section{Cluster states of qudits in the VBS formalism}\label{section_cluster_vbs}
Central to the workings of the one-way quantum computer (1WQC) is the cluster state \cite{cluster}.  Here we give a constructive definition of cluster states of qudits.
\begin{defn}
A cluster state consists of a lattice of qudits with some given neighbourhood scheme.  Each of the qudits on the lattice is individually prepared in the $\ket{+}$ state where
\be
\ket{+} = \frac{1}{\sqrt{d}} \sum_{j\in\Z_d}\ket{j}.
\ee
Then the two-qudit controlled-Z gate, $C_Z$ (as defined in equation \ref{defn_cz}), is applied once between each neighbouring pair of qudits.
\end{defn}

In this paper we will consider only linear and square lattices as they are sufficient for universal quantum computation.

\begin{figure}[htp]
  \begin{center}\includegraphics{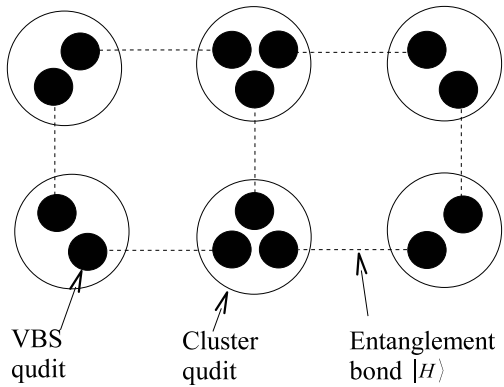}\end{center}
  \caption{An example VBS state}
  \label{figure_example_vbs}
\end{figure}

In the VBS formalism we describe the cluster state using a VBS state (generalising the procedure given for $d=2$ in \cite{vbs}).  A VBS state consists of pairs of qudits entangled in the state $\ket{H}$ as in figure \ref{figure_example_vbs}.
\be
\ket{H} = C_Z\ket{+}\ket{+} = \sum_{j,k \in \Z_d}\omega^{jk}\ket{j}\ket{k}.
\ee
Here we have ignored and shall continue to ignore normalisation factors throughout this paper.   Further properties of VBS states can be found in \cite{korepin1, korepin2}.

Given a particular cluster state we consider a corresponding VBS state with one pair of qudits entangled in the $\ket{H}$ state for each neighbouring pair of cluster qudits as shown in figure \ref{figure_example_vbs}.

We now show that the cluster state ``resides inside'' the corresponding VBS state, within the d-dimensional subspaces spanned by $\ket{j}...\ket{j}$ for $j \in \Z_d$ at each site of the VBS state.

\begin{thm}
For any VBS state introduce the projector
\be
\Pi_a = \sum_{j\in \Z_d}\tilde{\ket{j}}\bra{j}...\bra{j}
\ee
at each site $a$ where we have used a tilde to re-label the basis states after projection.  If we apply $\Pi_a$ for all $a$ to the VBS state we obtain (after re-normalising) the corresponding cluster state.
\end{thm}

\begin{figure}[htp]
  \begin{center}\includegraphics{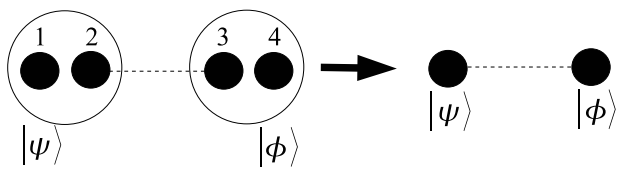}\end{center}
  \caption{Projecting one VBS bond to a cluster state}
  \label{fig_vbs_proj_1_dim}
\end{figure}

To prove the theorem we show that the combined action of the projector $\Pi_a$ at each site does indeed produce a cluster state on lattices of one and two dimensions.  Starting with the one-dimensional case we consider the lattice in figure \ref{fig_vbs_proj_1_dim}

Ignoring normalisation factors, the state on the left hand side of figure \ref{fig_vbs_proj_1_dim} is
\bea
\ket{\psi}_1\ket{H}_{23}\ket{\phi}_4 & = & \sum_j\psi_j\ket{j}_1 \sum_{mn}\omega^{mn}\ket{m}_2\ket{n}_3 \sum_k\phi_k\ket{k}_4.
\eea
The projector to apply is
\be
\sum_{p} \tilde{\ket{p}}\bra{p}_1\bra{p}_2 \otimes \sum_{q} \tilde{\ket{q}}\bra{q}_3\bra{q}_4.
\ee
This gives the state
\bea
\sum_{jkmnpq} \psi_j \phi_k \omega^{mn} \ip{p}{j} \ip{p}{m} \ip{q}{n} \ip{q}{k} \tilde{\ket{p}} \tilde{\ket{q}} & = & \sum_{pq} \psi_p \phi_q \omega^{pq} \tilde{\ket{p}} \tilde{\ket{q}} \\
& = & \label{eqn_vbs_proj_1_dim}C_Z\tilde{\ket{\psi}}\tilde{\ket{\phi}}.
\eea

\begin{figure}[htp]
  \begin{center}\includegraphics{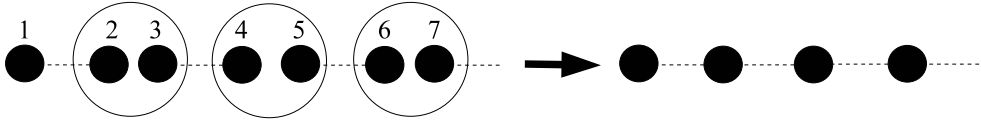}\end{center}
  \caption{Projecting to a cluster state on a one-dimensional lattice}
  \label{fig_vbs_proj_1_dim_gen}
\end{figure}

Using the derivation in equation \ref{eqn_vbs_proj_1_dim} we can extend this result to an arbitrary one-dimensional
cluster state. In figure \ref{fig_vbs_proj_1_dim_gen} the dashed lines already represent $C_Z \ket{+}\ket{+}$ and we
see that the projectors on sites $23$ and $45$ have the effect of applying a further $C_Z$ gate between the corresponding
cluster sites.  This is true as we continue down the lattice leaving a 1-dimensional cluster state.

\begin{figure}[htp]
  \begin{center}\includegraphics{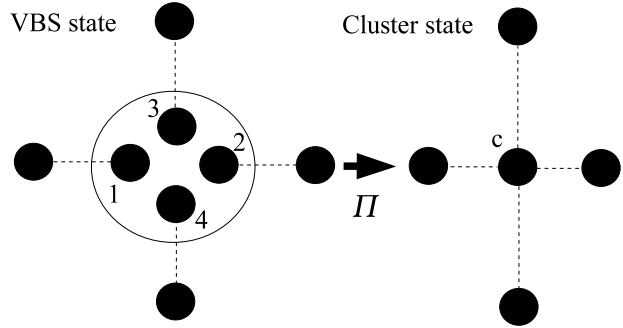}\end{center}
  \caption{A VBS state cluster state on a two dimensional lattice}
  \label{fig_vbs_proj_join}
\end{figure}

In the more general case of a two-dimensional cluster state we consider a VBS state site which represents a cluster qudit with four neighbours shown in figure \ref{fig_vbs_proj_join}.

The projector $\Pi$ for this cluster qudit is given by

\be
\Pi = \sum_k \tilde{\ket{k}}\bra{k}_1\bra{k}_2\bra{k}_3\bra{k}_4.
\ee
Which we can decompose into the sequential application of $\Pi_1$, $\Pi_2$ and $\Pi_3$ where
\begin{align}
\Pi_1 & = \sum_k \tilde{\ket{k}}_a \bra{k}_1\bra{k}_2,
& \Pi_2 & = \sum_k \tilde{\ket{k}}_b \bra{k}_a\bra{k}_3,
& \Pi_3 & = \sum_k \tilde{\ket{k}}_c \bra{k}_b\bra{k}_4.
\end{align}
Hence the result in equation \ref{eqn_vbs_proj_1_dim} applied successively shows that in general any two-dimensional VBS
state will project down to a cluster state completing the proof of the theorem.

\section{Teleportation-based quantum computation on VBS states}
In this section we describe how to perform universal quantum computation on VBS states using teleportation-based quantum computation (TQC) \cite{tqc_orig_nielsen, tqc_orig_leung}.  In section \ref{section_d1wqc} we show that this gives the functioning of the d-level one-way quantum computer (d1WQC) \cite{d1wqc} in the subspace that the cluster state resides in.

\subsection{Input and output qudits in the VBS formalism}
\begin{figure}[htp]
    \begin{center}
\includegraphics{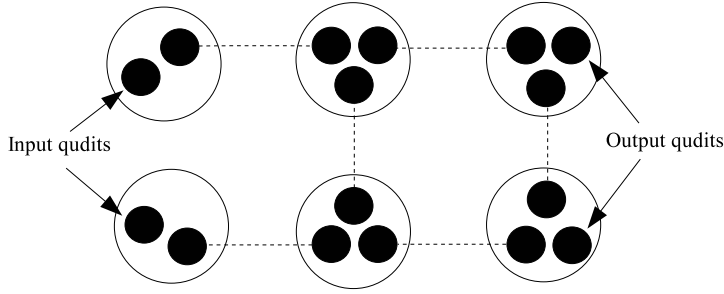}
    \end{center}
     \caption{Input and output in the VBS picture}
\label{vbs_with_io}

\end{figure}

Qudits that do not form part of a $\ket{H}$ bond are used for input and output.  The input qudits are placed as shown in figure \ref{vbs_with_io} in the desired state.  The output qudits are measured in the computational basis and these results are corrected using calculations from the other measurement results to form the classical output from the computation as we describe in the following. 

\subsection{Universality for quantum computation}
We show how to implement a universal set of gates on a VBS state.  The gates used are a parameterised one-qudit gate and the controlled-Z gate between two qudits.  As shown in the following these gates are implemented up to a random unitary error which is in the generalised Pauli group as described in section \ref{chapter_pauli}.  We show how these errors can be deterministically corrected for.

\subsubsection{A parameterised one-qudit gate}\label{dvbs_one_qudit}
In order to implement any one qudit gate we show that we implement any gate of a special form $U(\vec{c})$ that is parameterised by a vector $\vec{c} = (c_0=1,c_1,...,c_{d-1}) $ of $d$ complex numbers of modulus one.
The gate $U(\vec{c})$ is the defined as
\be \label{eqn_defn_uc}
U(\vec{c})\ket{j} = F \mbox{diag}(\vec{c}) = c_j \sum_{m \in \Z_d} \omega^{jm} \ket{m}.
\ee

\begin{figure}[htp]
    \begin{center}
\includegraphics{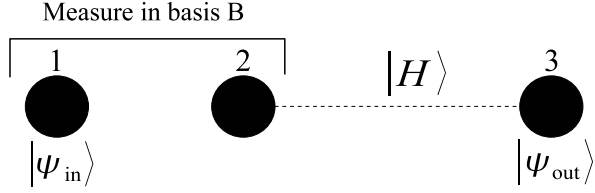}
    \end{center}
     \caption{Implementing the $U(\vec{c})$ gate modulo a random Pauli error}
  \label{vbs_1_qudit}   
\end{figure}

This particular set of gates is chosen because as shown in section \ref{section_uc_d1wqc} they have special properties in relation to the projectors $\Pi_a$.  $U(\vec{c})$ is implemented on a VBS state as a $d$-dimensional analog of the TQC as shown in figure \ref{vbs_1_qudit}.
Qudits $1$ and $2$ are measured in the basis $B$ defined as
\be\label{eqn_basis_uc}
B = \{\ket{\alpha_{st}} = (U(\vec{c})^\dagger X^sZ^t \otimes
I)\ket{H}\mbox{ : }s,t \in \Z_d \}.
\ee

We can see that the basis $B$ is a `twisted' generalised Bell basis and that so is equivalent to preparing qudit $1$ in the state $U(\vec{c})\ket{\psi}$ and measuring in the (untwisted) generalised Bell basis.  If the measurement result is $s,t \in \Z_d$ then we have teleported \cite{teleport, werner} the state $U(\vec{c})\ket{\psi}$ and qudit $3$ is left in the state
\be \label{eqn_post_uc}
Z^{-t}X^{-s}U(\vec{c}) \ket{\psi}.
\ee
Next we will show that the known Pauli error $Z^{-t}X^{-s}$ which is produced by the act of teleportation can be corrected for.

\subsubsection{Combining multiple $U(\vec{c})$ gates}\label{section_combine_uc}
We have shown how to implement the gate $U(\vec{c})$ modulo some random Pauli error $Z^tX^s$. Since $s$ and $t$ are known these errors can be tracked and corrected for. In order to do this when combining multiple $U(\vec{c})$ gates we will propagate all the errors to the end of the computation and correct for them last.  For this we need the  commutation relations of $U(\vec{c})$ with all Pauli errors.  It suffices to calculate them for the generators $X,Z$ of the Pauli group.

We introduce the following notation.  If $\vec{c} = (c_0,c_1,...,c_{d-1})$ then $\vec{c}_{++} = (c_1,c_2,...,c_{d-1},c_0)$. This gives
\be
U(\vec{c}_{++}) \ket{j} = c_{j+1} \sum_{m\in\Z_d}
\omega^{jm}\ket{m}.
\ee
Calculating the propagation for $Z$ and $U(\vec{c})$ we have
\be
U(\vec{c}) Z \ket{j} = U(\vec{c}) \omega^j \ket{j} = c_{j} \sum_{m\in\Z_d} \omega^{j(m+1)}\ket{m} =  X^{-1} U(\vec{c}) \ket{j}
\ee

Similarly for $X$ and $U(\vec{c})$ we have
\be
U(\vec{c})X \ket{j} =U(\vec{c}) \ket{j+1} = c_{j+1}\sum_{m} \omega^{m(j+1)}\ket{m} = Z U(\vec{c}_{++}) \ket{j}
\ee
Hence we have the propagation relations
\be
U(\vec{c}) Z  = X^{-1} U(\vec{c})
\ee
and
\be
U(\vec{c}) X  = Z U(\vec{c}_{++})
\ee
These relations allow us to implement many such gates of the form $U(\vec{c})$ and avoid having to correct the errors after each one by adapting the implementation of each gate depending upon the errors produced up to that point and tracking the errors for a future gate implementation.  This adds a requirement of performing steps of classical computation in between the measurement steps into the computational model.

At the end of the computation we restrict the output measurements to be in the computational basis.  Since each output qudit carries a Pauli error of $Z^tX^s$ if we obtain the measurement result $m \in Z_d$ we correct by taking the final result to be $m-s \modd$ since the $Z$ errors have no effect to measurements in the computational basis.

\subsubsection{Implementing $C_Z$}
\begin{figure}[htp]
    \begin{center}
\includegraphics{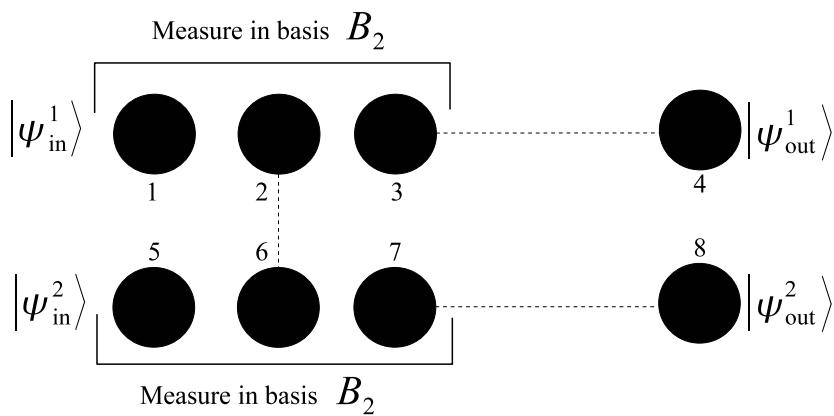}
    \end{center}
     \caption{Implementing a generalised $C_Z$ gate}
  \label{vbs_cz}   
\end{figure}

We implement the controlled-Z gate, $C_Z$,  on the VBS state as in figure \ref{vbs_cz}.  We use a three-qudit measurement in a basis $B_2$ where
\be\label{eqn_b2}
B_2 = \left\{X^r \otimes Z^s \otimes X^t \left( \sum_{m \in \Z_d}
\ket{m}\ket{m}\ket{m} \right) : r,s,t \in \Z_d \right\}
\ee

We denote a $C_Z$ gate applied between qudits $j$ and $k$ by $C_{Z(j,k)}$ where qudits $j$ and $k$ are the control and target qudits respectively.  The following lemma applies to figure \ref{vbs_cz}.
\begin{lem}\label{lemma_cz}
After measurement of qudits 1,2,3 in basis $B_2$ with measurement results $r,s,t \in \Z_d$ and measurement of qudits 5,6,7 in basis $B_2$ with results $u,v,w \in \Z_d$ the state of the subsystem consisting of qudits $4$ and $8$ is, up to a global phase, equal to
\be
Z_4^{t-r}Z_8^{w-u}X_4^{s+u}X_8^{v+r}F_4F_8C_{Z(4,8)}\ket{\psi_{in}^1}_4\ket{\psi_{in}^2}_8.
\ee
\end{lem}

This proof of this lemma is given in appendix B.

The implementation of $C_Z$, up to Pauli errors, is completed by applying the inverse Fourier transform, $F^\dagger$, to both output qudits using the techniques described in the previous section and theorem \ref{prop_univ_uc} below.

\subsubsection{Combining other gates with $C_Z$}
We have already seen the commutation relations of the one-qudit gate
$U(\vec{c})$ with the Pauli errors and how this allows us to correct
for all the errors at the end of the computation.  In the case of
$C_Z$ it is in the normaliser of the Pauli group with propagation
relations
\be
C_{Z(1,2)} Z_1 = Z_1 C_{Z(1,2)} \mbox{, }
C_{Z(1,2)}Z_2 = Z_2 C_{Z(1,2)}
\ee
and
\be
C_{Z(1,2)}X_1 = X_1Z_2C_{Z(1,2)}
\mbox{, }
C_{Z(1,2)} X_2 = Z_1X_2C_{Z(1,2)}.
\ee
From these relations we see that the implementation of the $C_Z$ gate is
not effected by Pauli error propagation.

\subsubsection{Proof of universality}
We now show that the ability to perform any gate of the form $U(\vec{c})$ and $C_Z$ allows for universal quantum computation.   Starting with the one-qudit case we have
\begin{thm}\label{prop_univ_uc}
Gates of the form $U(\vec{c})$ can be used to produce any one-qudit gate when $d$ is an odd prime
\end{thm}
\begin{proof}
Given a linearly independent set $\set{H_j}$ of $d^2$  hermitian matrices (each of size $(d \times d)$) we can write any unitary $U \in U(d)$ as
\be\label{eqn_herm_exp}
U = exp(i H) = exp\left(i \sum_{j = 1}^{d^2} \alpha_j H_j\right) = \prod_{j = 1}^{d^2} exp(i \beta_j H_j)
\ee
for some real parameters $\alpha_j, \beta_j$.

We now give such a set of linearly independent one-qudit Hermitian matrices and show how the corresponding one-parameter unitary gates can be implemented from gates of the form $U(\vec{c})$.

Let us choose $d + 1$ Pauli elements from $\pauliOne{d}$ such that the eigenvectors of these elements form a set of mutually unbiased bases \cite{pauli_mub}.  Let us denote these bases as
\be
\set{\ket{a_1}, \dots ,\ket{a_d}}, \set{\ket{b_1}, \dots ,\ket{b_d}}, \dots ,\set{\ket{e_1}, \dots , \ket{e_d}}.
\ee
We use these vectors to form a set of $d^2$ Hermitian operators
\be \label{eqn_linear_indep_herm}
\set{\ket{a_1}\bra{a_1}, \dots ,\ket{a_d}\bra{a_d}, \ket{b_2}\bra{b_2}, \dots ,\ket{b_d}\bra{b_d}, \ket{e_2}\bra{e_2}, \dots ,\ket{e_d}\bra{e_d}}
\ee
where we have omitted the first vector in all bases except the first basis.  We claim that this set is linearly independent. Since for any set of real numbers $\set{\alpha_j, \beta_k, \epsilon_k}_{j \in \set{1,\dots,d}, k \in \set{2, \dots , d}}$ if we have
\be
\alpha_1 \ket{a_1}\bra{a_1} + \dots + \alpha_d \ket{a_d}\bra{a_d} +   \beta_2 \ket{b_2}\bra{b_2} + \dots + \beta_d \ket{b_d}\bra{b_d}+  \epsilon_2 \ket{e_2}\bra{e_2}+ \dots + \epsilon_d \ket{e_d}\bra{e_d} = 0
\ee
then by applying $\bra{a_j} \dots \ket{a_j}$ to each side of the equation for each value of $j \in \set{1,\dots ,d}$ we obtain
\be
\alpha_j + \frac{1}{\sqrt{d}} \left( \beta_2 + \dots \beta_d + \epsilon_2+ \dots +\epsilon_d \right) = 0.
\ee
From this we conclude that all values of $\alpha_j$ are equal to $\alpha$, say.  Similarly we can argue that all values of $\beta_j$ ( $\dots ,\epsilon_j$) are equal to $\beta$ (respectively $\dots ,\epsilon$).  Let us now rewrite equation \ref{eqn_linear_indep_herm} as
\be
\alpha \left( \ket{a_1}\bra{a_1}+ \dots \ket{a_d}\bra{a_d} \right)+\beta \left( \ket{b_2}\bra{b_2}+ \dots \ket{b_d}\bra{b_d} \right) + \epsilon \left( \ket{e_2}\bra{e_2}+ \dots \ket{e_d}\bra{e_d} \right) = 0.
\ee
Then
\be
\alpha I + \beta \left( I - \ket{b_1}\bra{b_1} \right) + \dots + \epsilon \left( I - \ket{e_1}\bra{e_1} \right) = 0.
\ee
Rearranging we have
\be \label{eqn_streams}
\left( \alpha + \beta + \dots + \epsilon \right)I = \beta \ket{b_1}\bra{b_1} + \dots + \epsilon \ket{e_1}\bra{e_1}.
\ee
Applying $\bra{b_1} \dots \ket{b_1}$ to both sides gives
\be
\left( \alpha + \beta + \dots + \epsilon \right) =  \beta + \frac{1}{\sqrt{d}}\left(\gamma + \dots + \epsilon \right)
\ee
and applying $\bra{b_2} \dots \ket{b_2}$ to both sides gives
\be
\left( \alpha + \beta + \dots + \epsilon  \right) = \frac{1}{\sqrt{d}}\left( \gamma +\dots + \epsilon \right)
\ee
from which we conclude that $\beta = 0$ and we can similarly argue that $\gamma, \dots ,\epsilon = 0$.  Finally by applying $\bra{a_1} \dots \ket{a_1}$ to both side of equation \ref{eqn_streams} we obtain $\alpha = 0$ hence the set in equation \ref{eqn_linear_indep_herm} are linearly independent.

We now show that for each of the Hermitian operators $H$ given in equation \ref{eqn_linear_indep_herm} we can implement $U = exp(i\theta H)$ for any $\theta \in \R$ by gates of the form $U(\vec{c})$.

We first note that if $c_j = 1$ for all $j$ then $U(\vec{c})$ is the quantum Fourier transform $F$.  We can also can construct any diagonal matrix $D(\vec{c}) = diag(c_0,...,c_{d-1})$ as follows
\be
D(\vec{c}) = F^\dagger U(\vec{c}) = F^3 U(\vec{c}).
\ee
We can use this to construct the Clifford gate $S$
\be
S = D(c_j = \omega^{\frac{j(j+1)}{2}})
\ee

and also arbitrary rotations of the form
\be
exp(i \theta \ket{j}\bra{j}) = D(c_0,\dots, exp(i\theta \omega^{j}),\dots,c_{d-1}).
\ee

Then for any arbitrary element $P \in \pauliOne{d}$ we have (as is shown in appendix A) $P = C Z C^\dagger$ for some $C \in \cliffordOne{d}$ and $C$ is some product of $F$ and $S$ so $C$ can be expressed in terms of $U(\vec{c})$.  Then if $\ket{\lambda}$ is an eigenvector of $P$ then for some $j \in \Z_d$ we have up to a phase $\ket{\lambda} = C\ket{j}$.  It follows that
\bea
exp(i \theta \ket{\lambda}\bra{\lambda}) & = & exp(i \theta  C \ket{j}\bra{j}C^\dagger) \\
& = & C exp(i \theta \ket{j}\bra{j}) C^\dagger.
\eea
\end{proof}

\begin{cor}
Gates of the form $U(\vec{c})$ and $C_Z$ are universal for d-level quantum computation
\end{cor}
\begin{proof}
In \cite{universal_imprimitive} it is shown that any entangling two-qudit gate together with all one-qudit gates provides exact universality on an arbitrary number of qudits.  The authors show that the generalised controlled-Z gate $C_Z$ is entangling and hence, by theorem \ref{prop_univ_uc}, gates of the form $U(\vec{c})$ and $C_Z$ are universal for d-level quantum computation.
\end{proof}
We remark that we can have an approximately universal gate set of $\set{C_Z, F, D}$ where we have chosen $D$ to be a diagonal matrix where each entry is an irrational phase and each pair of phases differ by an irrational factor.

\section{The d1WQC in the VBS formalism}\label{section_d1wqc}
We saw in section \ref{section_cluster_vbs} that we can produce a cluster state by applying a projector of the form $
\Pi = \sum_k \tilde{\ket{k}}\bra{k}...\bra{k}$ to all the VBS qudits at each cluster site to give the cluster qudits.  In this section we show that the implementations of gate $U(\vec{c})$ and $C_Z$ given in the last section are well aligned with this projector: upon projection, these gate implementations are naturally converted into a pattern of one-qudit measurements on the cluster state thus deriving the measurement schemes for this set of universal gates on the d1WQC.

\subsection{Performing $U(\vec{c})$ in the d1WQC}\label{section_uc_d1wqc}
Considering the two-qudit measurement basis we used in section \ref{dvbs_one_qudit} to implement the $U(\vec{c})$ gate
\be
B = \{\ket{\alpha_{st}} = (U(\vec{c})^\dagger X^sZ^t \otimes I)\ket{H}\mbox{ : }s,t \in \Z_d \}.
\ee
We observe that the post measurement states corresponding to the measurement result $t=0$ all lie in the subspace of the projected cluster state since
\bea
(U(\vec{c})^\dagger X^s \otimes I)\ket{H} & = & (Z^sU(\vec{c})^\dagger  \otimes I)C_Z \ket{+}\ket{+} \\
& = & (Z^{-s} \otimes I) \sum_{jkl} \bar{c}_l \omega^{j(k-l)} \ket{l}\ket{k} \\
& = & \label{eqn_u_d1wqc}\sum_{l} \omega^{-sl} \bar{c}_l \ket{l} \ket{l}.
\eea

\begin{figure}[htp]
    \begin{center}
\includegraphics{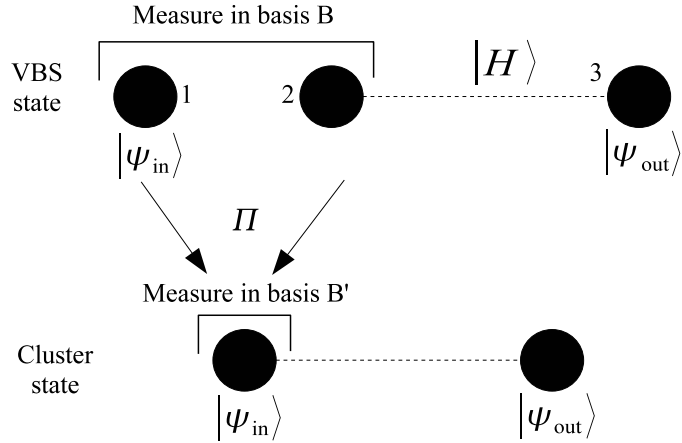}
    \end{center}
     \caption{Implementing $U(\vec{c})$ on the d1WQC}
  \label{fig_proj_uc}   
\end{figure}

Since $B$ is an orthogonal basis all the post measurement states corresponding to $t \neq 0$ lie in the orthogonal complement to this subspace.  So from equation \ref{eqn_post_uc} we see that in the restriction to the cluster state obtained by the projector $\Pi = \sum_{m} \ket{\tilde{m}} \bra{m} \bra{m}$ the final state of the second qudit is $X^{-s}U(\vec{c})$.  From equation \ref{eqn_u_d1wqc}, the one-qudit basis $B^\prime$ as shown in figure \ref{fig_proj_uc} on the d1WQC is thus
\be
B^\prime = \set{\sum_{l} \omega^{-sl} \bar{c}_l \tilde{\ket{l}} }_{s \in \Z_d} = \set{U^\dagger(\vec{c})\tilde{\ket{s}}}_{s \in Z_d}
\ee

Hence the effect of $\Pi$ is to restrict the outcomes of the $B$ measurement to have $t=0$, which is equivalent to performing a one-qudit measurement in basis $B'$ on the cluster state qudit.

By combining these measurements along a one dimensional cluster state and adaptively altering the basis to propagate Pauli errors to the end of the computation we can implement any one-qudit operation on the d1WQC.  Next we will see how to implement the two-qudit gate controlled-Z.

\subsection{Performing $C_Z$ in the d1WQC}
\begin{figure}[htp]
    \begin{center}
\includegraphics{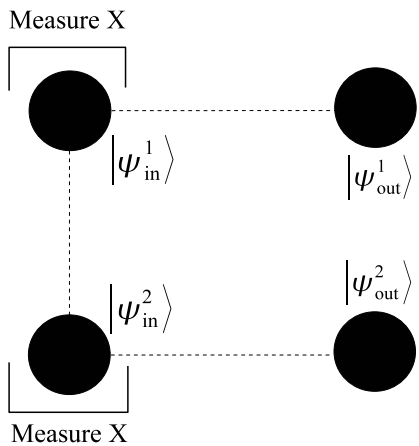}
    \end{center}
     \caption{Implementing a generalised $C_Z$ gate on the d1WQC}
  \label{fig_cz_d1wqc}   
\end{figure}

If we project the VBS state in figure \ref{vbs_cz}, which is used to implement the $C_Z$ gate, down to a cluster state we will obtain the state shown in figure \ref{fig_cz_d1wqc}.  Furthermore it is clear that the elements of the basis $B_2$ defined in equation \ref{eqn_b2} for which $r,t=0$ lie in the projected cluster state and the elements $r,t \neq 0$ lie in its orthogonal complement.  We consider the action of both measurements in the basis $B_2$ with corresponding measurement results $r,s,t$ and $u,v,w$. When restricted to $r,t,u,w=0$, which corresponds to the action on the projected cluster state shown in figure \ref{vbs_cz}, this produces, by lemma \ref{lemma_cz}, the output
\be\label{eqn_cz_d1wqc}
\ket{\psi_{out}^1}_1\ket{\psi_{out}^2}_2 = X_1^{s}X_2^{v}F_1F_2C_{Z(1,2)}\ket{\psi_{in}^1}_1\ket{\psi_{in}^2}_2.
\ee

Thus the measurement scheme to implement $C_Z$ on the d1WQC is obtained by applying the usual projector, $\Pi$, to the basis $B_2$.  Let $\ket{b}$ be an arbitrary basis vector corresponding to the measurement result $r,t=0$ such that $\ket{b} = (I \otimes Z^s \otimes I)\sum_m \ket{m}\ket{m}\ket{m}$ then
\begin{align}
\Pi \ket{b} &= \sum_k \tilde{\ket{k}} \bra{k}\bra{k}\bra{k} (I \otimes Z^s \otimes I)\sum_m \ket{m}\ket{m}\ket{m}\\
&=  \sum_{km}\omega^{sm}\tilde{\ket{k}} \bra{k}\bra{k}\bra{k} \ket{m}\ket{m}\ket{m} \\
&= \sum_k \omega^{sk} \tilde{\ket{k}}.
\end{align}
The scheme to implement $C_Z$ on the d1WQC, as shown in figure \ref{vbs_cz}, is to measure the two input qudits in the basis $\left\{F\ket{s}\right\}_{s \in Z_d}$ which is a measurement in the $X$ basis.  We must then implement the inverse Fourier transform $F^\dagger$ on both qudits and by equation \ref{eqn_cz_d1wqc} we will have implemented $C_Z$ up to a known Pauli error.

\section{Parallel complexity of d1WQC and extensions of the model}
We see from the construction of the d1WQC, in section \ref{section_combine_uc} that the only adaptations of measurements that we have to make are when we propagate the teleportation errors from the Pauli group through the gates of the form $U(\vec{c})$.  In the case where we wish to implement one-qudit Clifford gates, we may leave our implementation involving $U(\vec{c})$ gate unchanged and calculate the propagation of the Pauli errors though the Clifford gate.  In this way if the circuit we wish to implement is a Clifford circuit then we may apply all the one-qudit measurements on the d1WQC in parallel.

The VBS formalism that we have described provides a fundamental connection between d1WQC and the process of teleportation and this relationship leads to a wide class of natural extensions and generalisations of 1WQC.  Werner \cite{werner} has shown that there exists a wide variety of inequivalent teleportation schemes in dimensions greater than 2.  For example any set of operators that form a unitary operator basis may be used to construct a teleportation scheme.  Furthermore it can be shown that even in dimension 2 there exist still more possible teleportation schemes in which the Bell measurement is replaced by a POVM \cite{peres}.

Any of these teleportation schemes may then be used in a VBS setting resulting in new classes of measurement-based models of quantum computation.  In each such formalism we have a set of ``teleportation correction operators'' analogous to the Pauli operations in standard teleportation, and an associated normaliser group.  Circuits of the latter operators would then lead to further new classes of parallelisable quantum algorithms.  These issues will be developed in a later paper.

\section{Conclusion}
We have shown how to interpret the workings of d-level one-way quantum computation in terms of d-level valence bond solids.   We constructed cluster states of qudits using this formalism and derived implementations of a universal set of gates on the d1WQC using one-qudit measurements.  We also showed that, analogously to the qubit case, the set of circuits in the Clifford group, $\clifford{d}{n}$, can be implemented in one parallel time step of quantum measurements on the d1WQC followed by some classical computation and we have characterised the structure of the Clifford group for spaces of prime dimension.

\section*{Acknowledgements}
I would like to thank Richard Jozsa for much advice and help during the production of this paper and Ashley Montanaro, Noah Linden, Andreas Winter, Tobias Osborne and David Fattal for helpful discussions.   This work was supported by the UK Engineering and Physical Sciences Research Council QIP-IRC grant and the U.K. Government Communications Head Quarters.

Note added in proof: after this paper was completed we noticed the appearance of \cite{hall} which treats some of the same issues from a different perspective.

\renewcommand{\theequation}{A.\arabic{equation}}
\setcounter{equation}{0}  
\section*{Appendix A: Generating the Clifford group for d-level systems where d is prime}\label{chap_dcliff}
\label{chapter_d_gen_cliff}

In this appendix we fully characterise the Clifford group $\clifford{d}{n}$, for the case where $d$ is an odd prime, by showing that all its elements can be generated, up to a global phase factor, by circuits consisting of $C_X$, $F$ and $S$ as defined in section \ref{chapter_pauli}.

We note the following commutation properties
\be
ZX= \omega XZ \mbox{ and }\label{one_qudit_comm_rel}(Z^aX^b)(Z^cX^d) = \omega^{ad-bc}(Z^cX^d)(Z^aX^b).
\ee

If we write $P = (Z_1^{a_1}X_1^{b_1}\dots Z_n^{a_n}X_n^{b_n})$ and $Q =
(Z_1^{c_1}X_1^{d_1}\dots Z_n^{c_n}X_n^{d_n})$ then
\be\label{eqn_pauli_comm_rel}
PQ=\omega^{\sum_{i=1}^na_id_i-b_ic_i}QP = \omega^{(P,Q)}QP
\ee
where we use the following notation
\be
(P,Q) = \sum_{i=1}^na_id_i-b_ic_i.
\ee

The generalised Clifford group $\clifford{d}{n}$ on $n$ qudits is defined in definition \ref{defn_gen_clifford} as the normaliser of $\pauli{d}{n}$.  Each $C \in \clifford{d}{n}$ induces an endomorphism of
$\pauli{d}{n}$ by its action under conjugation.  We write
\be
P \mapsto_C Q\mbox{ for }P,Q \in \pauli{d}{n}\mbox{ when }CPC^{-1} = Q.
\ee
Sometimes it will be useful to consider two elements $P,Q \in \pauli{d}{n}$ as equivalent if they differ only by a global phase.  In this way we can represent each member  $c Z_1^{a_1}X_1^{b_1}\dots Z_n^{a_n}X_n^{b_n} \in \pauli{d}{n}$ (where $c$ is a phase) up to global phase as:
\be
(a_1,b_1,\dots ,a_n,b_n) \in \Z_d^{2n}\label{eqn_pauli_no_phase}.
\ee

In view of the commutation relation \ref{eqn_pauli_comm_rel}, products in $\pauli{d}{n}$  correspond up to a phase to addition of the corresponding vectors in $\Z_{d}^{2n}$.  Furthermore the action of Clifford operations is linear: if we use elements in $\set{Z_1, X_1,\dots ,Z_n,X_n}$, where $Z_i$ is the $n$-qudit operator which acts as $Z$ on qubit $i$ and the identity elsewhere, as a basis of $\pauli{d}{n}$ we can represent the action of $C$ up to a global phase as a $2n \times 2n$ matrix $M(C)$ with entries in $\Z_d$.

$F$ and $S$ induce the following mappings on $\pauliOne{d}$
\begin{align}\label{eqn_prop_qft}
& X \mapsto_F Z\mbox{ and } Z \mapsto_{F} X^{-1} \\
& Z \mapsto_S Z \mbox{ and }X \mapsto_S ZX
\end{align}

so the matrix representations $M(F)$ and $M(S)$ are
\begin{align}\label{eqn_ms}
& M(F) \left( \begin{smallmatrix} a \\ b \end{smallmatrix} \right) =
\left( \begin{smallmatrix} 0 & 1 \\ -1 & 0 \end{smallmatrix}
\right)\left( \begin{smallmatrix} a \\ b \end{smallmatrix} \right) =
\left( \begin{smallmatrix} b \\ -a \end{smallmatrix} \right) \\
& \label{eqn_mf}
M(S)\left( \begin{smallmatrix} a \\ b \end{smallmatrix} \right) =
\left( \begin{smallmatrix} 1 & 1 \\ 0 & 1 \end{smallmatrix}
\right)\left( \begin{smallmatrix} a \\ b \end{smallmatrix} \right) =
\left( \begin{smallmatrix} a+b \\ b \end{smallmatrix} \right).
\end{align}

\begin{lem}
$F^{-1}$, $S^{-1}$, $Z$, $X$, $Z^{-1}$ and $X^{-1}$ can all be constructed from $\set{F,S}$.
\end{lem}
\begin{proof}
Firstly we note that $F^4 = I$ so $F^{-1} = F^3$ and $S^d = I$ so $S^{-1} = S^{d-1}$.  Then we have
\be
Z = F^2S^{-1}F^2S.
\ee
Then since $Z^d = I$ we have $Z^{-1} = Z^{d-1}$.  We can use this to construct $X$ since
\be
X = FZ^{-1} F^{-1}.
\ee
Finally we have $X^d = I$ so $X^{-1} = X^{d-1}$.
\end{proof}

$C_X \in \clifford{d}{2}$ can be seen from the following mappings on $\pauli{d}{2}$
\be
Z_1 \mapsto_{C_X} Z_1\mbox{, }X_1 \mapsto_{C_X} X_1X_2\mbox{, }Z_2
\mapsto_{C_X} Z_1^{-1}Z_2\mbox{, }X_2 \mapsto_{C_X} X_2.
\ee

Similarly $C_Z\in \clifford{d}{2}$ since
\be\label{eqn_prop_cz}
Z_1 \mapsto_{C_Z} Z_1\mbox{, }X_1 \mapsto_{C_Z} X_1Z_2\mbox{, }Z_2
\mapsto_{C_Z} Z_2\mbox{, }X_2 \mapsto_{C_Z} Z_1X_2.
\ee

Since the normalising property of actions is preserved by composition and tensor product we see that any gate that can be constructed (in the quantum circuit sense \cite{nielsen_chuang}) from gates in the Clifford group must itself be in the Clifford group.

\begin{lem}{\label{lem_constr_Z}}
$C_Z$ can be constructed from $\{C_X,F\}$
\end{lem}
\begin{proof}
\be
C_{Z(1,2)} = F_2C_{X(1,2)}F^{-1}_2.
\ee
\end{proof}

\begin{defn}
An arbitrary controlled Pauli operator $C_{X^{s}Z^t}$ with $s,t \in
\Z_d$ is defined as
\begin{eqnarray}{\label{defn_controlled_pauli}}
C_{X^{s}Z^t} \ket{j}\ket{k}& = & \ket{j}(X^{s}Z^t)^j\ket{k} \\
           & = & \omega^{\frac{stj(j-1)}{2} + tjk}\ket{j}\ket{k + sj}.
\end{eqnarray}
\end{defn}
This then produces the following mappings
\be \label{eqn_map_gen_control}
Z_1 \mapsto_{C_{X^sZ^t}} Z_1\mbox{, }X_1 \mapsto_{C_{X^sZ^t}}
X_1X_2^{s}Z_2^t\mbox{, }Z_2 \mapsto_{C_{X^sZ^t}} Z_1^{-s}Z_2\mbox{,
}X_2 \mapsto_{C_{X^sZ^t}} Z_1^tX_2.
\ee

\begin{lem}{\label{constr_arb_pauli}}
$C_{X^sZ^t}$ can be constructed from $\{C_X,F,S\}$ if the qudit dimension $d$ is an odd integer.
\end{lem}
Remark: If $d$ is an even integer then the definition of $S$ needs to be modified in order for it to be a valid Clifford operation, and then this lemma remains valid.
\begin{proof}
We have already seen in equation \ref{defn_controlled_pauli} that
\be
C_{X^sZ^t}\ket{j}\ket{k} =
\omega^{\frac{stj(j-1)}{2}}(C_X)^s(C_Z)^t\ket{j}\ket{k}
\ee
where $C_Z$ is suitably constructed (by lemma \ref{lem_constr_Z}).  We note that since $d$ is an odd integer
\begin{align}
S\ket{j} & = \omega^{\frac{j(j+1)}{2}}\ket{j} \\
SZ^{-1}\ket{j} & = \omega^{\frac{j(j-1)}{2}}\ket{j} \\
C_{X^sZ^t} & = (C_X)^s(C_Z)^t(SZ^{-1})_1^{st}.
\end{align}
\end{proof}
\begin{defn}
The $SWAP$ gate is defined as $SWAP\ket{j}\ket{k} = \ket{k}\ket{j}$.
\end{defn}
\begin{lem}
$SWAP$ can be constructed from $\{C_X, F\}$.
\end{lem}
\begin{proof}
We can construct a $C_X$ gate that uses the second qudit as control and the first as target
\be
C_{X(2,1)} = F_1F_2^{-1}C_{X(1,2)}F_1^{-1}F_2.
\ee
Then $SWAP$ is constructed using the following identity
\be
SWAP = C_{X(1,2)}C_{X(2,1)}^{-1}C_{X(1,2)}F_2^2
\ee
where, since $C_{X(2,1)}^d = I \otimes I$ we have $C_{X(2,1)}^{-1} = C_{X(2,1)}^{d-1}$.
\end{proof}
The construction of the $SWAP$ gate from the gate set $\{C_X,F,S\}$ allows constructions in which multiple qudit
gates can be applied to non-local qudits.  Often the quantum circuit model allows for non-local applications of two-qudit gates.  The above lemma shows that such an assumption is not necessary for the construction of the Clifford group.

Now we turn our attention to associations defined on subsets of the the Pauli group.
\begin{defn}{\label{comm_rel}}
Let $\set{P_i}$ and $\set{\bar{P_i}}$ be any subsets of $\pauli{d}{n}$ of the same size.  We say that the association $P_i \mapsto \bar{P_i}$ is commutation relation preserving (CRP) if
\be
(P_i,P_j) = (\bar{P_i},\bar{P_j}) \mbox{ for all }i,j.
\ee
\end{defn}

\begin{lem}{\label{cliff_is_crp}}
The maps induced by conjugation with Clifford group operations are CRP on $\pauli{d}{n}$.
\end{lem}
\begin{proof}
For any $P, Q \in \pauli{d}{n}$ we have $PQ = \omega^{(P,Q)} QP$ so
for $U \in \clifford{d}{n}$ we have
\begin{equation}
(UPU^{-1})(UQU^{-1}) = U(PQ)U^{-1} = \omega^{(P,Q)}
U(QP)U^{-1} =\omega^{(P,Q)} (UQU^{-1})(UPU^{-1}).
\end{equation}
\end{proof}

\begin{lem}{\label{lem_one_gen}}
Given any CRP association $A$ of one-qudit Pauli operators $Z \mapsto
Z^aX^b$ and $X \mapsto Z^cX^d$ we can construct an operator from $
\{F, S\}$ whose action generates this association.
\end{lem}
\begin{proof}
In terms of the representation of equation \ref{eqn_pauli_no_phase}, the matrix of the association $A$ is
\be
M(A) = \left(\begin{smallmatrix} a & c \\ b & d \end{smallmatrix} \right).
\ee
From equation \ref{one_qudit_comm_rel} and the fact that $A$ is CRP
we can deduce that
\be
ad-bc = 1
\ee
so $M(A) \in SL(2,\Z_d)$.   In \cite{lang} it is shown that the matrices $M(S)$ and $M(F)$ in equations \ref{eqn_ms} and \ref{eqn_mf} generate $SL(2,\Z_d)$. Hence any such $M(A)$ can be generated by $M(F)$ and $M(S)$.
\end{proof}

Having established this result for CRP associations defined on $\pauliOne{d}$ we now extend this to $\pauli{d}{n}$.  An outline of the remainder of the proof is as follows:  In lemma \ref{det_one_row} we show that given  $P,Q \in \pauli{d}{n}$ such that $P = Z_1^{a_1}X_1^{b_1}\dots Z_n^{a_n}X_n^{b_n}$ and $Q = Z_1^{c_1}X_1^{d_1}\dots Z_n^{c_n}X_n^{d_n}$ we may assume wlog (modulo some suitable mapping constructed from $\set{C_X,F,S}$) that there exists some $j \in \set{1,\dots ,n}$ such that $a_jd_j - b_jc_j = 1$.  In lemma \ref{lem_first_qudit} we show that if $(P,Q) = 1$ then we can assume wlog (modulo some mapping constructed from $\set{C_X,F,S}$) that $P  = X \otimes P^\prime$ and $Q = Z \otimes Q^\prime$.   Then we take some arbitrary CRP association $X_i \mapsto \bar{X_i}$ and $Z_i \mapsto \bar{Z_i}$  and the main part of the proof is to establish that it may be constructed from $\set{C_X,F,S}$.  In lemma \ref{lem_build_u} we take such a CRP association and assume that $\bar{X}_1 = X \otimes P^\prime$ and $\bar{Z}_1 = Z \otimes Q^\prime$ and construct a gate $U$ from $\set{C_X,F,S}$ such that $X_1 \mapsto_U X \otimes P^\prime$ and $Z_1 \mapsto_U Z \otimes Q^\prime$.  Taking this gate $U$ we show in lemma \ref{lem_rs} that there exists $R_i, S_i \in \pauli{d}{n-1}$ such that $I \otimes R_i \mapsto_U \bar{X}_i\mbox{ and }I \otimes S_i \mapsto_U \bar{Z}_i$.  Using the Pauli elements $R_i$ and $S_i$ we show in lemma \ref{lem_v} that the $n-1$ qudit association $V$ defined by $X_i \mapsto_V I \otimes R_{i}$ and $Z_i \mapsto_V I \otimes S_{i}$ is CRP.  This leads to lemma \ref{lem_combine_maps} in which we show that the arbitrary association $X_i \mapsto \bar{X}_i$ and $Z_i \mapsto \bar{Z}_i$ is satisfied by the mapping induced by $U(I\otimes V)$.  Using the preceding results we proceed by induction in  theorem \ref{main_result} to show that any such CRP association can be constructed from $\set{C_X,F,S}$ and in corollary \ref{main_theorem} that $\clifford{d}{n}$ is generated by $\set{C_X,F,S}$ when $d$ is prime.

\begin{lem}{\label{det_one_row}}
Given $P,Q \in \pauli{d}{n}$ such that $(P,Q) = 1$ and
\be
P = Z_1^{a_1}X_1^{b_1}\dots Z_n^{a_n}X_n^{b_n}\mbox{, }Q = Z_1^{c_1}X_1^{d_1}\dots Z_n^{c_n}X_n^{d_n}
\ee
there exists a construction $M$ from $\{C_X,F,S\}$ such that
\be
P \mapsto_M Z_1^{a^\prime _1}X_1^{b^\prime _1}\dots Z_n^{a^\prime _n}X_n^{b^\prime _n}\mbox{ and }Q \mapsto_M Z_1^{c^\prime_1}X_1^{d^\prime _1}\dots Z_n^{c^\prime_n}X_n^{^\prime _n}
\ee
and there exists $j \in \{1,\dots ,n\}$ with
$a^\prime_jd^\prime_j - b^\prime_jc^\prime_j = 1$.
\end{lem}
\begin{proof}

Since $(P,Q) = 1$ we have
\be{\label{eqn_det_sum}}
\sum_{i=1}^na_id_i - b_ic_i = 1
\ee
so we can choose $j$ such that
\be{\label{eqn_j_row}}
a_jd_j - b_jc_j \ne 0.
\ee
If $a_jd_j - b_jc_j = 1$ then the mapping $M$ is trivial and the
proof completes. Otherwise there must exist $k \ne j$ such that
\be{\label{eqn_k_row}}
a_kd_k - b_kc_k \ne 0.
\ee
The construction for $M$ follows. Firstly if $b_j \ne 0$ we take $g$ such that $a_j + gb_j = 0$ (the existence of such a $g$ following from $d$ being prime) and apply $FS^g$ to $P$ and $Q$ by conjugation to the $j^{th}$ qudit.  This maps $P$ to $\bar{P}$, say, where $\bar{P}$ is of the form such that $\bar{b_j} = 0$.  Given this mapping let us assume that the original $P$ was of the form such that
\be\label{eqn_bj}
b_j = 0.
\ee

We apply by conjugation a $C_{X^sZ^t}$ gate to $P$ and $Q$ with the $j^{th}$ qudit as control
and $k^{th}$ qudit as target.  Using the relations given in equation \ref{eqn_map_gen_control} we obtain
\begin{align}
& a^\prime_j = a_j - sa_k + tb_k \mbox{, } \mbox{ } b^\prime_j  = b_j \\
& c^\prime_j = c_j - sc_k + td_k \mbox{, } \mbox{ } d^\prime_j  = d_j.
\end{align}
Hence given $b_j = 0$ we have
\be
a^\prime_jd^\prime_j - b^\prime_jc^\prime_j = (a_j - sa_k+tb_k)d_j.
\ee
We observe from equation \ref{eqn_k_row} that $a_k$ and $b_k$
can not both be zero and since $d_j \ne 0$ (by equations \ref{eqn_j_row} and
\ref{eqn_bj}) we can choose $s,t \in \Z_d$ such that
\be
(a_j - sa_k+tb_k)d_j = 1
\ee
by the fact that $d$ is prime.  Hence we have $a^\prime_jd^\prime_j - b^\prime_jc^\prime_j  = 1$ as desired.
\end{proof}

\begin{lem}{\label{lem_first_qudit}}
Given $P,Q \in \pauli{d}{n}$ such that $(P,Q) = 1$ there is a
construction $W$ from $\{C_X,F,S\}$ such that $P \mapsto_W X
\otimes P^\prime$ and $Q \mapsto_W Z \otimes Q^\prime$ for some
$P^\prime, Q^\prime \in \pauli{d}{n-1}$.
\end{lem}
\begin{proof}
Using the same notation as, and by an application of, lemma
\ref{det_one_row} we assume wlog that for some $j \in \set{1,\dots ,n}$
\be
a_jd_j - b_jc_j = 1.
\ee

We construct $W$ by performing a $SWAP$ between the $1^{st}$ and $j^{th}$ qudits followed by a one-qudit mapping induced by $L$ on the $1^{st}$ qudit where the matrix of $L$ is
\be
M(L) = \left(\begin{matrix} d_j & -c_j \\
-b_j & a_j \end{matrix} \right).
\ee
$L$ produces the desired mapping since
\be
Z^{a_j}X^{b_j} \mapsto_L X \mbox{, } Z^{c_j}X^{d_j }\mapsto_L Z.
\ee
Furthermore since $M(L)$ has unit determinant it can be constructed
from $\{F, S\}$ by lemma \ref{lem_one_gen}.
\end{proof}

\begin{lem}{\label{lem_build_u}}
Suppose we have a CRP association $X_i \mapsto \bar{X}_i$ and $Z_i \mapsto \bar{Z}_i$ defined on $\pauli{d}{n}$ where $d$ is an odd prime and let us assume wlog (by lemma \ref{lem_first_qudit}) that $\bar{X}_1 = X \otimes P^\prime$ and $\bar{Z}_1 = Z \otimes Q^\prime$ with $P^\prime,Q^\prime \in \pauli{d}{n-1}$. Then there exists a construction $U$ from $\{C_X,F,S\}$ such that $X_1 \mapsto_U X \otimes P^\prime$ and $Z_1 \mapsto_U Z \otimes Q^\prime$.
\end{lem}
\begin{proof}
Let us write
\begin{align}
& X \otimes P^\prime =Z_1^{a_1}X_1^{b_1} \dots Z_n^{a_n}X_n^{b_n} \\
& Z \otimes Q^\prime =Z_1^{c_1}X_1^{d_1} \dots Z_n^{c_n}X_n^{d_n}
\end{align}
We construct a circuit $P^{\prime}_{impl}$ from $P^\prime$ in which we perform $C_{X^{b_i}Z^{a_i}}$ between the $1^{st}$ and the $i^{th}$ qudit (using the $1^{st}$ as control) for each $i \in \{2,\dots ,n\}$. We repeat the construction of $P^{\prime}_{impl}$ with the indices of $Q^{\prime}$ to produce $Q^{\prime}_{impl}$.  The construction of $U$ is then
\begin{equation}
 U = F_1  Q^{\prime}_{impl} F^{-1}_1 P^{\prime}_{impl}.
\end{equation}

We now justify this construction.  The image of $Z_1$ by $U$ can be seen to be $Z \otimes Q^{\prime}$ from the following sequence of mappings:
\be
Z_1 \mapsto_{P^\prime_{impl}} Z_1 \mapsto_{F_{1}^{-1}} X_1 \mapsto_{Q^{\prime}_{impl}} X \otimes Q^{\prime} \mapsto_{F_1} Z \otimes Q^{\prime}
\ee
where we have used the mappings in \ref{eqn_map_gen_control} to deduce that $Z_1$ commutes with each $C_{X^{b_i}Z^{a_i}(1,i)}$ from $P^\prime_{impl}$ and the image of $X_1$ under conjugation with each $C_{X^{d_i}Z^{c_i}(1,i)}$ from $Q^\prime_{impl}$ is $X_1Z_i^{c_i}X_i^{d_i}$.

Now let us look at the image of $X_1$ when conjugated by $U$ we have
\be
X_1 \mapsto_{P^\prime_{impl}} X \otimes P^\prime \mapsto_{F_{1}^{-1}} Z^{-1}_1 \otimes P^\prime.
\ee

Then $Z^{-1}_1$ commutes with $Q^{\prime}_{impl}$ and is mapped to $X_1$ by the final $F_1$.
We must consider the image of the elements of $P^{\prime}$ by $Q^{\prime}_{impl}$.  The image of $Z_{i}^{a_i}X_{i}^{b_i}$ on the target qudit under the action of $C_{X_{i}^{d_i}Z_i^{a_i}}$ is $Z^{b_ic_i - a_id_i}$ on the control and $Z^{a_i}X^{b_i}$ on the target and the target is as desired.  The contribution to the power of $Z$ on the control by the image of $P^{\prime}$ by $Q^{\prime}_{impl}$ is then
\be
\sum^{n}_{i=2}b_ic_i-a_id_i.
\ee
Since the mapping is CRP we have
\be
1=(X_1,Z_1)=(X \otimes P^\prime, Z \otimes Q^\prime) = \sum_{i=1}^n a_id_i - b_ic_i.
\ee
Furthermore $a_1d_1 - b_1c_1 = 1$ so $\sum^{n}_{i=2}b_ic_i-a_id_i = 0$ and hence $X_1 \mapsto_U X \otimes P^\prime$ as desired.
\end{proof}

\begin{lem}{\label{lem_rs}}
The Clifford circuit $U$ in lemma \ref{lem_build_u} has the property
that
\be
I \otimes R_i \mapsto_U \bar{X}_i\mbox{ and }I \otimes S_i \mapsto_U
\bar{Z}_i
\ee
for some $R_i,S_i \in \pauli{d}{n-1}$ and all $i \in \{2,\dots ,n\}$.
\end{lem}
\begin{proof}
Since $U$ is a Clifford operation so is $U^{-1}$.  We have the CRP map $X_i \mapsto \bar{X}_i \mapsto_{U^{-1}} \bar{X}_i^\prime$ and $Z_i \mapsto \bar{Z}_i \mapsto_{U^{-1}} \bar{Z}_i^\prime$. For $i \in \{2,\dots ,n\}$  $\bar{X}_i^\prime$ commutes with both $X_1$ and $Z_1$ and so is of the form $I \otimes R_i$. Similarly $\bar{Z}_i^\prime$ commutes with both $X_1$ and $Z_1$ and is of the form $I \otimes S_i$.
\end{proof}

\begin{lem}{\label{lem_v}}
The $n-1$ qudit association $V$ (acting on qudits $2$ to $n$) given by $X_i \mapsto_V I \otimes R_{i}$ and $Z_i \mapsto_V I \otimes S_{i}$ for $i \in
\{2,\dots ,n\}$ is CRP.
\end{lem}
\begin{proof}
Since $X_i \mapsto \bar{X}_i \mapsto_{U^{-1}} I
\otimes R_i$ and $Z_i \mapsto \bar{Z}_i \mapsto_{U^{-1}} I
\otimes S_i$ is CRP we have for $i,j \in \{2,\dots ,n\}$
\be
(S_i,S_j) = (X_i,X_j) = 0 \mbox{, } (R_i,R_j) = (Z_i,Z_j) = 0 \mbox{, } (S_i,R_j) = (X_i,X_j) = \delta_{ij}.
\ee
Hence V is CRP.
\end{proof}
\begin{lem}{\label{lem_combine_maps}}
The mapping induced by $U(I \otimes V)$  where $U$ and $V$ are defined in lemmas \ref{lem_build_u} and
\ref{lem_v} is such that
\be
X_i \mapsto_{U(I \otimes V)} \bar{X}_i\mbox{ and }Z_i \mapsto_{U(I
\otimes V)} \bar{Z}_i
\ee
\end{lem}
\begin{proof}
The result follows from:
\be
X_1 \mapsto_{I \otimes V} X_1 \mapsto_U X \otimes P^\prime = \bar{X_1} \mbox{, } Z_1 \mapsto_{I \otimes V} Z_1 \mapsto_U Z \otimes Q^\prime = \bar{Z_1}
\ee
and
\be
X_i \mapsto_{I \otimes V} I \otimes R_i \mapsto_U \bar{X_i} \mbox{, } Z_i \mapsto_{I \otimes V} I \otimes S_i \mapsto_U \bar{Z_i}\mbox{ for }i \in \{2,\dots ,n\}.
\ee
\end{proof}
\begin{thm}{\label{main_result}}
Any CRP association $X_i \mapsto \bar{X_i}$ and $Z_i \mapsto \bar{Z_i}$ for $i \in \set{1, \dots , n}$ defined on $\pauli{d}{n}$ where $d$
is an odd prime can be constructed from $ \{C_X, F, S \} $.
\end{thm}
\begin{proof}
We proceed by induction on $n$ where the base of $n=1$ is provided by lemma \ref{lem_one_gen}.  We assume that any CRP association on $(n-1)$ qudits can be constructed from $\{C_X, F, S\}$.  For the $n$-qudit CRP association $X_i \mapsto \bar{X_i}$, $Z_i \mapsto \bar{Z_i}$
there exists, by lemma \ref{lem_first_qudit}, a construction $W$ from $\{C_X, F, S\}$ such that $\bar{X_1} \mapsto_W X \otimes P^\prime$ and $\bar{Z_1} \mapsto_W Z \otimes Q^\prime$.  Suppose $W$ maps
$\bar{X_i} \mapsto \bar{X^\prime_i}$ and $\bar{Z_i} \mapsto
\bar{Z^\prime_i}$. By lemmas \ref{lem_build_u}, \ref{lem_v} and
\ref{lem_combine_maps} there exists a CRP map $U(I \otimes V)$ which
maps $X_i \mapsto \bar{X^\prime_i}$ and $Z_i \mapsto
\bar{Z^\prime_i}$.  So $W^{-1} U(I \otimes V)$ maps $X_i \mapsto
\bar{X}_i$ and $Z_i \mapsto \bar{Z}_i$.  $U$ has a construction from
$\{C_X, F, S\}$ by \ref{lem_build_u} and since V acts on $n-1$
qudits there exists a construction for it from $\{C_X, F, S\}$ by
the inductive hypothesis.
\end{proof}
\begin{cor}{\label{main_theorem}}
The Clifford group on $n$ qudits is generated by $ \{C_X, F, S \}
$ when the dimension $d$ of a qudit is an odd prime.
\end{cor}
\begin{proof}{ }
Any Clifford group mapping is fully defined by its action on $X_i$ and $Z_i$. Furthermore, by lemma \ref{cliff_is_crp} this association is CRP, so by theorem \ref{main_result} it can be constructed from $ \{C_X, F, S \} $.
\end{proof}

\renewcommand{\theequation}{B.\arabic{equation}}
\setcounter{equation}{0}  
\section*{Appendix B: Proof of lemma \ref{lemma_cz}} \label{section_proof_lem_impl_cz}
\begin{proof}
The state of the system in figure \ref{vbs_cz} before measurement is
\beas
\ket{VBS} & = &\ket{\psi^1_{in}}_1\ket{\psi^2_{in}}_5\ket{H}_{26}\ket{H}_{34}\ket{H}_{78}
\\
& = & \left(\sum_a \psi^1_a\ket{a}_1 \right) \left(\sum_b
\psi^2_b\ket{b}_5 \right) \left(\sum_{c,d}
\omega^{cd}\ket{c}_2\ket{d}_6 \right) \left(\sum_{e,f}
\omega^{ef}\ket{e}_3\ket{f}_4 \right) \left(\sum_{g,h}
\omega^{gh}\ket{g}_7\ket{h}_8 \right) \\
& = & \sum_{abcdefgh} \psi^1_a \psi^2_b \omega^{cd+ef+gh}
\ket{a}_1\ket{c}_2\ket{e}_3\ket{f}_4\ket{b}_5\ket{d}_6\ket{g}_7\ket{h}_8
\eeas
If the measurement results are $r,s,t,u,v,w \in \Z_d$ then the following projector is applied to the VBS state
\be
\sum_{mnpq}\omega^{s(m-n)+v(p-q)} \ket{m+r}_1 \ket{m}_2 \ket{m+t}_3 \ket{p+u}_5 \ket{p}_6 \ket{p+w}_7 \bra{n+s}_1 \bra{n}_2 \bra{n+t}_3 \bra{q+u}_5 \bra{q}_6 \bra{q+w}_7
\ee
Applying $\Pi_{B_2}$ to $\ket{VBS}$ we get six indices removed with
the following relations
\be
a = n+r\mbox{, }
b  =  q+u \mbox{, }
c  =  n \mbox{, }
d  =  q \mbox{, }
e  =  n+t \mbox{, }
g  =  q+w
\ee
giving
\bea
\Pi_{B_2} \ket{VBS} & = &
\sum_{mnpqfh}\psi^1_{n+r}\psi^2_{q+u}\omega^{s(m-n)+v(p-q)+nq+nf+tf+qh+wh}\\ &
&
\ket{m+r}_1\ket{m}_2\ket{m+t}_3\ket{f}_4\ket{p+u}_5\ket{p}_6\ket{p+w}_7\ket{h}_8
\\
& = & \label{eqn_out_cz}
\left(\sum_{nqfh}\psi^1_{n+r}\psi^2_{q+u}\omega^{-sn-vq+nq+nf+tf+qh+wh}\ket{f}_4\ket{h}_8
\right) \\
& & \mbox{ }\mbox{ } \otimes \left( ... \right)_{123567}
\eea

The restriction to qudits 4 and 8 of this state is recognised with a simple calculation as
\be
Z_4^tZ_8^wF_4F_8C_{Z(4,8)}Z_4^{-s}Z_8^{-v}X_4^{-r}X_8^{-u} \ket{\psi_{in}^1}_4 \ket{\psi_{in}^2}_8.
\ee
We can propagate all the Pauli terms to the left hand side of the expression using the propagation  relations given in equations \ref{eqn_prop_qft} and \ref{eqn_prop_cz} so that equating up to a global phase
\be
Z_4^tZ_8^wF_4F_8C_{Z(4,8)}Z_4^{-s}Z_8^{-v}X_4^{-r}X_8^{-u} = Z_4^{t-r}Z_8^{w-u}X_4^{s+u}X_8^{v+r}F_4F_8C_{Z(4,8)}
\ee

\end{proof}


\bibliographystyle{unsrt}
\bibliography{d-vbs}
\end{document}